\documentclass[aip,pof,reprint]{revtex4-1}

\usepackage{graphicx}
\usepackage{amssymb}
\usepackage{wasysym}
\usepackage{color}

\newcommand{\rme}{\mathrm{e}}
\newcommand{\avg}[1]{\langle {#1} \rangle}
\newcommand{\sR}{\mathcal{R}}
\newcommand{\sP}{\mathcal{P}}

\begin{document}
\title{Prediction of turbulence control for arbitrary periodic spanwise wall movement}
\author{Andrea Cimarelli} \affiliation{DIEM, Universit\`a di Bologna, Via Fontanelle 40, 47121 Forl\`i, Italy}
\author{Bettina Frohnapfel} \affiliation{Institute of Fluid Mechanics, Karlsruhe Institute of Technology, Kaiserstr. 10, 76131 Karlsruhe, Germany}
\affiliation{Center of Smart Interfaces, TU Darmstadt, Petersenstr. 17, 64287 Darmstadt, Germany}
\author{Yosuke Hasegawa}
\affiliation{Institute of Industrial Science, The University of Tokyo, 4-6-1 Komaba, Meguro-ku, Tokyo 153-8505, Japan}
\author{Elisabetta De Angelis} \affiliation{DIEM, Universit\`a di Bologna, Via Fontanelle 40, 47121 Forl\`i, Italy}
\author{Maurizio Quadrio} \affiliation{Dept. Aerospace Sciences and Technologies, Politecnico di Milano, Via La Masa 34, 20156 Milano, Italy}
\email{maurizio.quadrio@polimi.it}

\begin{abstract}

In order to generalize the well-known spanwise-oscillating-wall technique for drag reduction, non-sinusoidal oscillations of a solid wall are considered as a means to alter the skin-friction drag in a turbulent channel flow. A series of Direct Numerical Simulations is conducted to evaluate the control performance of nine different temporal waveforms, in addition to the usual sinusoid, systematically changing the wave amplitude and the period for each waveform.

The turbulent average spanwise motion is found to coincide with the laminar Stokes solution that is constructed, for the generic waveform, through harmonic superposition. This allows us to define and compute, for each waveform, a new penetration depth of the Stokes layer which correlates with the amount of turbulent drag reduction, and eventually to predict both turbulent drag reduction and net energy saving rate for arbitrary waveforms.

Among the waveforms considered, the maximum net energy saving rate is obtained by the sinusoidal wave at its optimal amplitude and period. However, the sinusoid is not the best waveform at every point in the parameter space. Our predictive tool offers simple guidelines to design waveforms that outperform the sinusoid for given (suboptimal) amplitude and period of oscillation. This is potentially interesting in view of applications, where physical limitations often preclude the actuator to reach its optimal operating conditions.

\end{abstract}

\maketitle


\section{Introduction}

The efficient use of energy in systems where a relative motion between a solid wall and a fluid takes place is perhaps the most important driving factor that supports the research effort into aerodynamic drag reduction. A significant body of literature is devoted to devising and testing methods for reducing the turbulent skin-friction drag, which arises from the interaction between turbulence and a wall even in the most simplified geometry (plane channel flow), and is also a key source of aerodynamic drag in many important applications.

Several existing control strategies belong to the class of active methods, that are located midway between the interestingly simple passive methods, e.g. riblets \cite{bechert-bartenwerfer-1989}, and the interestingly effective closed-loop (feedback) control methods \cite{kim-bewley-2007}. Unfortunately, the effectiveness of the former is still too limited to justify their deployment in applications, whereas the latter definitely lack simplicity, since the required large number of miniaturized sensors and actuators still prevents their near-future employment \cite{kasagi-suzuki-fukagata-2009}. Within a third and intermediate category, other drag reduction techniques, referred to as open-loop, provide higher drag reduction than existing passive methods while being less complex than feedback-control methods. In particular, open-loop techniques that rely on the spanwise forcing of the near-wall turbulent flow have been shown to yield large reduction of friction drag and interestingly positive energy budgets in numerical simulations \cite{quadrio-2011}, and first laboratory experiments have already been carried out \cite{auteri-etal-2010, gouder-potter-morrison-2013, choi-jukes-whalley-2011}. The present paper deals with the  spanwise oscillating-wall technique, i.e. the cyclic movement of the wall in the spanwise direction \cite{jung-mangiavacchi-akhavan-1992}. Although providing limited performance, especially in terms of net energy savings, the sinusoidal wall oscillation is representative of a larger group \cite{quadrio-2011}, and is one of the simplest available techniques, since only two control parameters exist, i.e. the oscillation period and the maximum wall velocity during the cycle (also referred to as oscillation amplitude).

Most existing open-loop control strategies assume a sinusoidal waveform for a control input without a compelling theoretical basis. When attempting to verify these control strategies in experiments, however, various constraints are inevitably placed on the properties of the control input by the employed actuators, and the waveform is often non-sinusoidal by practical necessity. For example, the streamwise-traveling-wave in the Milano pipe experiment \cite{auteri-etal-2010} had to deal with a streamwise distribution of the spanwise velocity which was not sinusoidal, but piecewise-constant. The experimental realization of the oscillating wall concept by Gouder {\em et al.} \cite{gouder-potter-morrison-2013} was limited by the maximum displacement achievable by their actuator. The plasma actuators by K.-S.Choi \cite{choi-jukes-whalley-2011} produced a localized forcing that was just an approximation of the sinusoidal distribution originally employed in the numerical simulations \cite{du-karniadakis-2000}. Hence, it is of particular importance to identify the optimal waveform to achieve best control performance among those realizable with a certain actuator.

Quadrio \& Ricco \cite{quadrio-ricco-2004} were the first to mention that, in flow control with wall oscillation, a third parameter besides oscillation period and amplitude enters the picture, i.e. the maximum displacement of the wall during the oscillation cycle. If the wall oscillates sinusoidally in time, however, only two of these three parameters are independent, and for example the maximum displacement can be deduced once period and amplitude are known. In the present paper, we intend to truly open up the third dimension in the parameters space, by investigating the behavior of the oscillating wall when the temporal waveform is considered as a free parameter, and the constraint of sinusoidal oscillation is lifted. Given the highly non-linear process that in principle links the oscillation of the wall and the reduction of the turbulent drag, the control performance for arbitrary waveforms cannot be predicted based on the present knowledge. As a starting point, we select a set of waveforms and study, via several numerical experiments, how the drag-reduction and energetic performances of the oscillating wall depend on the waveform as well as on the oscillation amplitude and period.
In this process, we opt for comparing different periodic waveforms at the same oscillation amplitude and period. Although different criteria could have been employed (e.g. considering waveforms with the same period and displacement, or with the same power input), these two parameters characterize in a simple way the main features of a generic waveform and play a first-order role in determining the control performance.

Guided by our numerical experiments, in this paper we aim at obtaining results of general validity, so that a predictive tool for the control performance of non-sinusoidal wall oscillations can eventually be developed. In this process, we take advantage of the laminar solution that exists for the spanwise flow alone: the Stokes oscillating boundary layer. This laminar solution has been shown \cite{quadrio-sibilla-2000} to describe the average spanwise turbulent flow for sinusoidal wall oscillations and has been used \cite{quadrio-ricco-2004} to interpret some turbulent drag reduction properties of the oscillating wall. Quadrio and Ricco \cite{quadrio-ricco-2011} extended this solution to the case of streamwise-traveling waves, and introduced the analytical expression of the so-called Generalized Stokes oscillating layer. Here, we attempt a complementary generalization to account for a generic (periodic) temporal waveform.

\section{The numerical experiments}
\label{sec:numerics}

The performance of non-sinusoidal spanwise wall oscillations is assessed using Direct Numerical Simulation (DNS) of a turbulent channel flow. We employ a computer code \cite{luchini-quadrio-2006} that solves the incompressible Navier--Stokes equations written in terms of wall-normal velocity and vorticity. Discretization is spectral (dealiased Fourier) in the homogeneous $x$ and $z$ directions, whereas compact, explicit fourth-order finite differences are used in the wall-normal $y$ direction. Time advancement is carried out with a partially implicit, third-order Crank--Nicholson/Runge--Kutta scheme.
The Reynolds number of the reference simulation without wall oscillation is $Re =  U_b h / \nu = 3173$, where $U_b$ is the bulk velocity, $h$ is half the channel gap and $\nu$ is  the kinematic viscosity of the fluid. This corresponds to a value of the friction-based Reynolds number $Re_\tau=200$. Unless otherwise specified, $h$ and $U_b$ are chosen as length and velocity scales. The computational domain is $9.6 \times 2 \times 4.8 $ along $x$, $y$ and $z$ directions, with a number of modes (before dealiasing) or grid points of $192 \times 128 \times 192$ respectively, which yields a spatial resolution (for the reference case) of $\Delta x^+ = 10$ and $\Delta z^+ = 5$; wall-normal resolution varies from $\Delta y^+ \approx 0.4$ near the wall to $\Delta y^+ \approx 5.4$ at the centerline (the superscript $+$ as customary implies non-dimensionalization in viscous units, where the friction velocity $u_\tau$ of the reference flow is used). Every simulation is started from the same initial condition of fully developed turbulent channel flow with stationary wall. When the wall moves, drag reduction takes place: since the flow rate is kept constant, the space-averaged streamwise pressure gradient and the friction drag decrease. The total integration time for each simulation is $95$ wash-out time units, corresponding to $10,000$ viscous time units. After the beginning of the oscillating movement of the walls at $t=0$, a certain time interval is needed for the flow to reach the new equilibrium state. Therefore, the initial 25\% of the time integration interval is not considered in the time average procedure.

\begin{figure}
\centering
\includegraphics[width=0.75\textwidth]{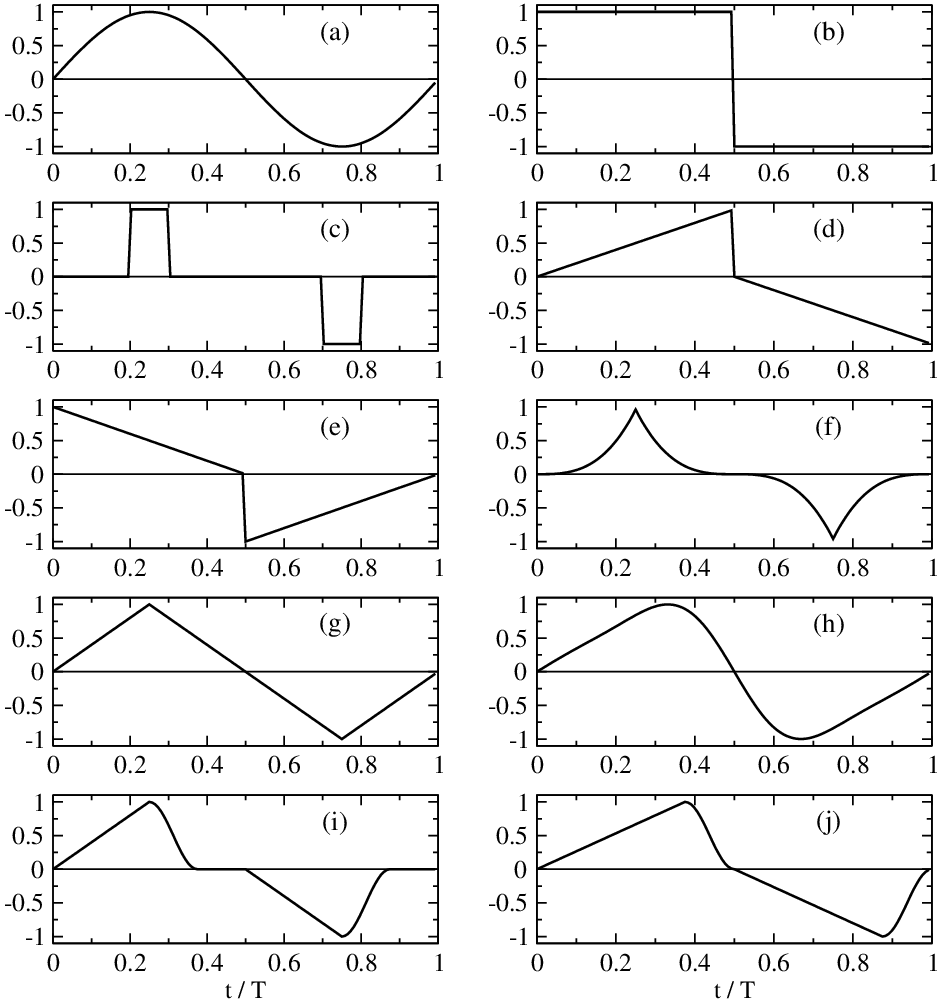}
\caption{Temporal waveforms of the considered wall oscillation.
The following symbols are used throughout the paper for the different waveforms: (a) ($\circ$), (b) ($\blacklozenge$),
(c) ($\ast$), (d) ($+$), (e) ($\blacksquare$),
(f) ($\CIRCLE$), (g) ($\blacktriangleleft$), (h) ($\blacktriangleright$), (i) ($\blacktriangle$) and (j) ($\blacktriangledown$).}
\label{fig:waveforms}
\end{figure}

\subsection{The waveforms}

Various waveforms of the temporal oscillation of the wall are examined. We consider spanwise wall velocities $W_w(t)$ varying in time as
\begin{equation}
W_w(t) = W_m f_\alpha \left( \frac{t}{T} \right),
\end{equation}
where $f_\alpha$ (with $\alpha = a, \ldots , j$) are ten periodic functions of unit period with values ranging from $-1$ to $1$. All the considered oscillations thus have period $T$ and amplitude $W_m$.
The investigated waveforms $f_\alpha$, which include the sinusoid, are sketched in figure \ref{fig:waveforms}; they are representative of the different features which may characterize non-sinuosidal oscillations in practice, featuring discontinuities in velocity and acceleration profiles, large and small accelerations, phase shifts and different fractions of the period with constant velocity and even zero velocity.
For every waveform, the oscillation parameters $W_m$ and $T$ are varied around the optimal values $T^+_{opt}$ and $W^+_{m,opt}$ that yield the maximum net energy saving rate for the sinusoidal oscillation. This particular condition has been carefully determined by Quadrio \& Ricco \cite{quadrio-ricco-2004} and corresponds to $T^+_{opt}=125$ and $W_{m,opt}^+=4.5$. We consider a parametric set of variations from this basic case, by changing $W_m^+$ and $T^+$ to values twice and one half of the optimal value. The nine combinations of $W_m^+=2.25, 4.5, 9$ and $T^+=62.5, 125, 250$ combined with the 10 considered waveforms, give rise to a total of 90 DNS carried out and reported in this study.

\subsection{Performance indicators}

To analyze the performance of the oscillating wall as a drag reduction technique, we follow Kasagi et al. \cite{kasagi-hasegawa-fukagata-2009} and employ three dimensionless indicators ($R, P_{in}, S$) that describe the global energy budget. The time-averaged pumping power per unit channel wall area in the fixed-wall case is given by
\[
\sP_0 = \frac{U_b}{t_f- t_i} \int_{t_i}^{t_f} \avg{ \tau_x^{0} } dt
\]
where $\avg{ \tau_x^0 }$ is the space-averaged streamwise component of the wall shear stress for the reference case and $t_i$ and $t_f$ mark the time average interval. The reduction of pumping power in the oscillating wall case is thus
\[
\sR = \frac{U_b}{t_f- t_i} \int_{t_i}^{t_f} \left( \avg{\tau_x^0} - \avg{\tau_x} \right) dt ,
\]
where $\avg{\tau_x}$ is the mean wall shear stress in the oscillating case. The power required to move the walls, at least in an idealized system where mechanical losses of the actuation device are neglected,
reads
\[
\sP_{in} = \frac{1}{t_f- t_i} \int_{t_i}^{t_f} \avg{ W_w \tau_z} dt ,
\]
where $\tau_z$ is the spanwise component of the wall shear stress, and $W_w$ is the wall velocity.

The ratio of $\sR$ and $\sP_{in}$ and the pumping power of the uncontrolled reference case, $\sP_0$, yields the following dimensionless performance indicators:
\[
R = \sR / \sP_0 , \qquad P_{in} = \sP_{in} / \sP_0 .
\]

Let us point out that, for the present problem where DNS is run at constant flow rate, the numerical value of the reduction of the normalized pumping power equals the reduction of drag, so that in the following we will refer to $R$ in both ways.
A net energy saving rate, that accounts for the benefits as well as for the energy costs of the moving wall can be easily defined as $S = R - P_{in} .$

\section{Results: Performance in terms of $P_{in}$, $R$ and $S$}
\label{sec:results}

\begin{figure}
\centering
\includegraphics[width=0.8\textwidth]{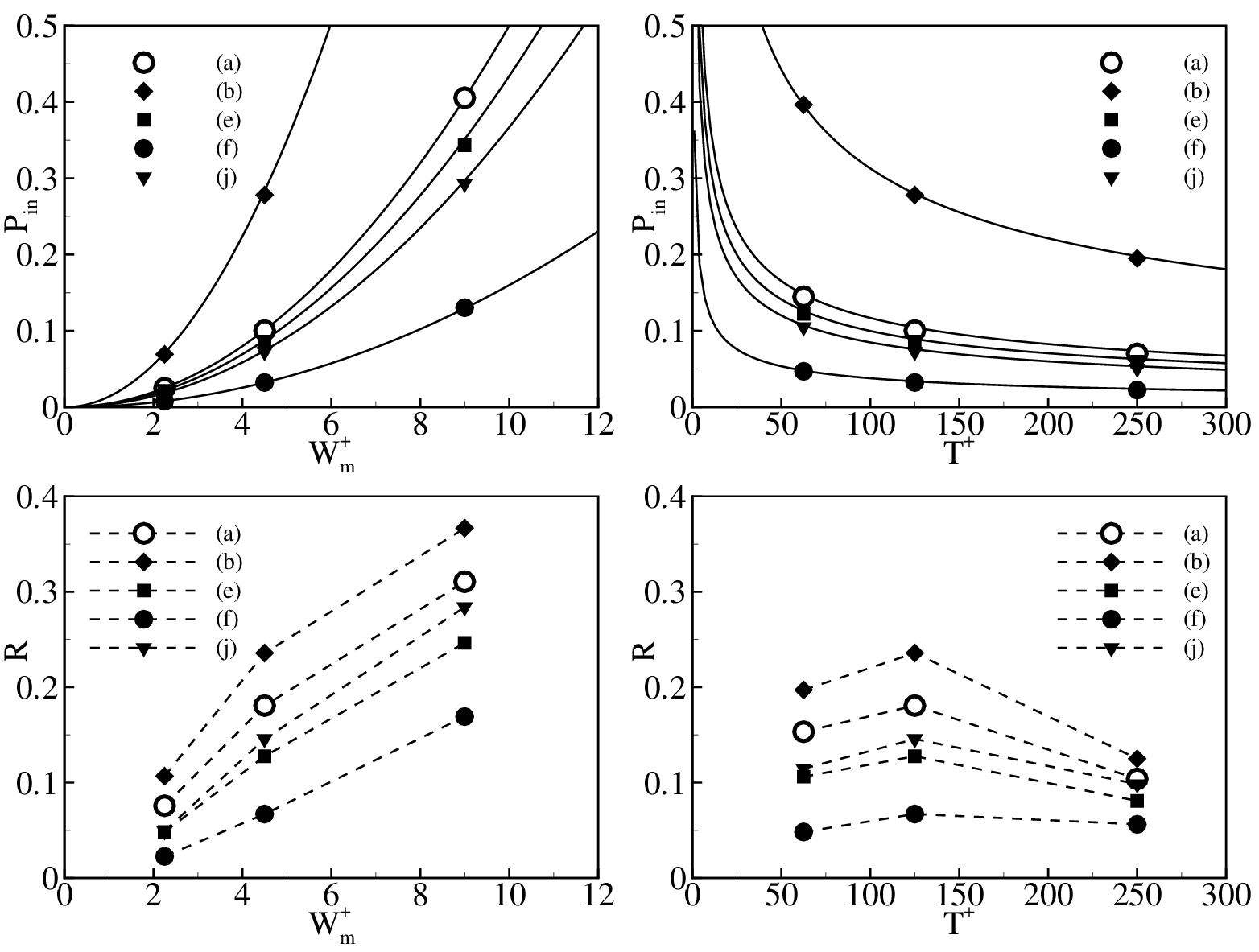}
\includegraphics[width=0.8\textwidth]{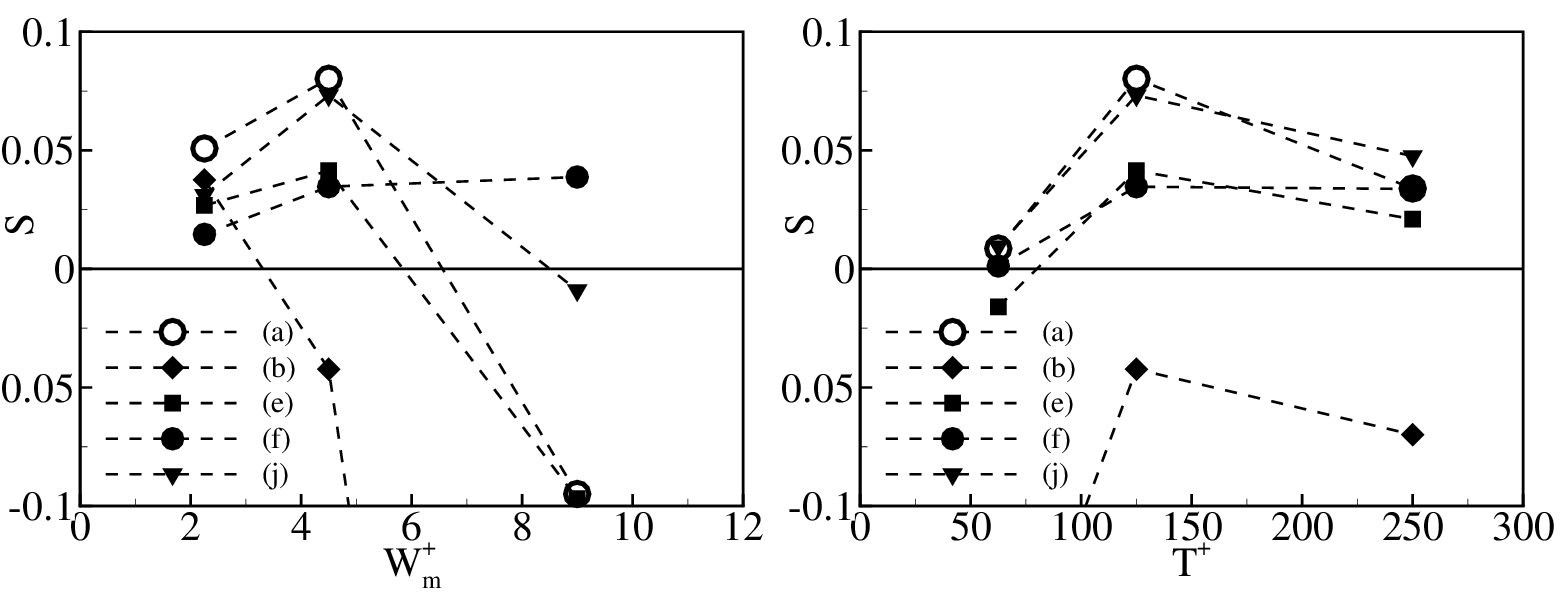}
\caption{Power consumption rate $P_{in}$ (top), reduction of pumping power $R$ (center), and net energy saving rate $S$ as a function of $W_m^+$ for $T^+ = 125$ (left) and of $T^+$ for $W_m^+ =4.5$ (right) for five selected waveforms, see figure \ref{fig:waveforms} for reference. The sinusoidal case is shown with the empty symbol. The continuous lines in the top plots are $P_{in} \propto W_m^2 \sqrt{\pi / Re T}$.}
\label{fig:Pin_R_WT}
\end{figure}

The power consumption $P_{in}$, the drag reduction rate $R$, and the net energy saving rate $S$ as functions of the oscillation period $T$, and maximum wall velocity $W_m$, for five representative waveforms are shown in figure \ref{fig:Pin_R_WT}. Notwithstanding the marked quantitative differences, both $P_{in}$ and $R$ qualitatively behave in the $(T,W_m)$-space like the sinusoidal case. The power consumption $P_{in}$ for each value of $T$ increases with $W_m$. For constant $W_m$, $P_{in}$ decreases with increasing $T$. The drag reduction rate, $R$, always presents its maximum at the intermediate period $T^+=125$, and it increases monotonically with increasing $W_m$. Overall, the specific waveform $f_\alpha$ enters the picture by affecting the quantitative values of $P_{in}$ and $R$.
All the considered non-sinusoidal waveforms at the optimal conditions $T_{opt}^+ = 125$ and $W_{m,opt}^+= 4.5$ yield a best $S$ which is smaller than $S=0.078$ obtained with sinusoidal oscillations. This optimum value is in agreement with what is reported in the literature \cite{quadrio-ricco-2004}. For nearly all cases considered the balance between $R$ and $P_{in}$ is such that $S>0$ at low $W_m$, whereas at larger $W_m$ higher values of $R$ but even larger values of $P_{in}$ are obtained, such that $S$ is reduced to negative values.

In general, most of the waveforms show a maximum of $S$ at the same intermediate values of $(T,W_m)$ as those of the sinusoidal case. An exception is given by waveform $(f)$, for which $S$ does not exhibit a local maximum in the investigated parameter space and still appears to increase at large $T$ and $W_m$ implying that the optimal oscillating period (in terms of max $S$) lies beyond $T^+ = 150$. A similar behavior is observed for waveform $(c)$ (not shown). Even though the corresponding $S$ here are very small or negative, the possibility of shifting the optimal oscillation period for net energy saving rate above $T^+= 150$ by properly selecting the waveform is an interesting possibility, as will be discussed later.
\begin{figure}
\centering
\includegraphics[width=0.4\textwidth]{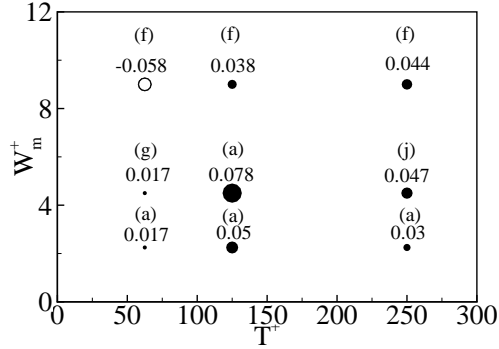}
\caption{Maximum values of net energy saving, $S$, for each pair of $(T,W_m)$-values investigated. Full circles correspond to $S>$, open circles to $S<0$ while the diameter of the circles reflects the obtained absolute value of $S$. The letter above each circle indicates the waveform yielding the best $S$.}
\label{fig:Gain}
\end{figure}

We emphasize that, even if the sinusoidal waveform is found to give the best overall $S$, regions of the $(T,W_m)$-space exist where non-sinusoidal wall oscillations locally yield values of $S$ that are higher than the sinusoidal one. This result is summarized in figure \ref{fig:Gain} where the best waveform (in terms of $S$) for each considered pair of $(T,W_m)$-values is shown. For example, waveform $(f)$ works well for large $W_m$. This waveform is characterized by low values for both $P_{in}$ and $R$. Since at large $W_m$ $S$ is more influenced by power consumption than by drag reduction, this results in a relatively high $S$. The possibility of a local increase of the net energy saving  above the value of the sinusoid is potentially important when working with the optimal oscillation parameters is not possible or impractical.

\section{Waveform generalization of the laminar Stokes solution}

The alternating boundary layer created by a sinusoidally oscillating wall in a quiescent fluid is described by the solution of the so-called Stokes second problem \cite{schlichting-gersten-2000}. It has been shown that this laminar solution also describes well the spanwise component of a turbulent channel or pipe flow modified by the sinusoidally oscillating wall, when properly averaged in space and considered as a function of the oscillation phase.
Since properties of the Stokes layer have been successfully related by several authors to $P_{in}$ and $R$, we intend to develop a predictive tool based upon the generalized form of the laminar Stokes solution.
In a first step we confirm that the agreement between the laminar solution $w_{St}(y,t)$ and the space-mean, phase-averaged spanwise turbulent profile $\avg{w}(y,\tau)$ holds for non-sinusoidal waveforms too.
The equivalence between $w_{St}(y,t)$ and $\avg{w}(y,\tau)$ implies that the spanwise space-averaged momentum equation reduces to a diffusion equation analogous to that of the laminar Stokes problem, i.e.
\begin{equation}
\frac{\partial \avg{w} }{\partial t} = \frac{1}{Re} \frac{\partial^2 \avg{w} }{\partial y^2} \, ,
\label{eq:Spanwise_averaged_eq}
\end{equation}
and the Reynolds stress term $\partial \langle w^\prime v^\prime \rangle / \partial y$ is negligible \cite{ricco-quadrio-2008}.
Since Eq.~(\ref{eq:Spanwise_averaged_eq}) is linear, the natural starting point is to consider a harmonic decomposition of the waveform, and to build the solution as linear superposition of the various Stokes components.
In the general case, the time-dependent boundary condition for the diffusion equation (\ref{eq:Spanwise_averaged_eq}) can be expressed via the following Fourier series
\begin{equation}
W_w(t) = W_m \sum_{n=1}^{+\infty} A_n \rme^{j (2 \pi n / T) t} + c.c.
\label{eq:Wall_decomposition}
\end{equation}
where $j$ is the imaginary unit, $A_n$ is the complex coefficient of the $n$-th  Fourier component and $c.c.$ stands for complex conjugate. The resulting expression for the waveform-generalized spanwise Stokes layer, obtained by superposition of the elementary solutions, reads as
\begin{equation}
w_{St}(y,t) = W_m \sum_{n=1}^{+\infty} A_n \rme^{-\sqrt{n} y / \delta}
\rme^{j \left[ \left (2 \pi n / T \right ) t - \sqrt{n} y / \delta \right ]} + c.c. ,
\label{eq:Generalized_Stokes_layer}
\end{equation}
where the wall-normal lengthscale $\delta$ is defined as
$\delta = \sqrt{{T}/{\pi Re}}$.

As shown in the next paragraph where equation (\ref{eq:Generalized_Stokes_layer}) is used for the prediction of $P_{in}$, the superposition of laminar solutions provides a robust description of the transverse boundary layer created by a generic wall movement. The knowledge of the spanwise velocity profile is exploited in what follows to derive a predictive tool for the control performance for wall oscillations of arbitrary waveform.

\subsection{A predictive tool}

The input power required by the sinusoidal oscillation is obtained from the Stokes solution and reads $\sP_{in} ={W_m^2}/{2} \sqrt{{\pi}/{T Re}}$. For a generic waveform, expressed through the Fourier series (\ref{eq:Wall_decomposition}), the same quantity becomes
\begin{equation}
\sP_{in} = W_m^2 \sqrt{\frac{\pi }{T Re}} \sum_{n=1}^{+\infty} 2 | A_n |^2 \sqrt{n} \, .
\label{eq:Generalized_Pc}
\end{equation}

\begin{figure}
\centering
\includegraphics[width=0.8\textwidth]{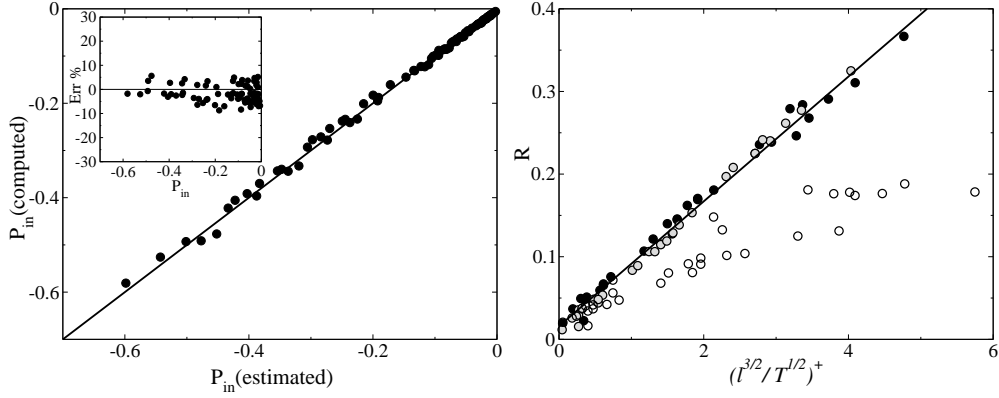}
\caption{
Left: power consumption data $P_{in}$ for all the simulations versus the analytical prediction based on the Stokes layer assumption, equation (\ref{eq:Generalized_Pc}). The inset plot shows the precentage error.
Right: turbulent drag reduction $R$ as a function of $(\ell^{3/2} T^{-1/2})^+$, with $\ell$ computed from Eq. (\ref{eq:Stokes_rms2}) using a threshold $\sigma_{th}^+ = 0.85$ corresponding to $W_{th}=1.2$. Gray symbols are for $T^+=62.5$, black symbols for $T^+=125$ and open symbols for $T^+=250$.
}
\label{fig:prediction}
\end{figure}

As shown in Fig. \ref{fig:prediction} (left), the power consumption computed with Eq.~(\ref{eq:Generalized_Pc}) is in excellent agreement with the simulation results for the entire dataset. The inset highlights how the percentage error remains small even when the absolute value of $P_{in}$ approaches zero. Eq.~(\ref{eq:Generalized_Pc}) can thus be used to predict $P_{in}$ for arbitrary values of $T$ and $W_m$, as well as for arbitrary waveforms. Moreover, the same equation highlights that $\sP_{in} \propto W_m^2 \sqrt{\pi / Re T}$ such that the qualitative dependency of $P_{in}$ on $T$ and $W_m$ is independent of the waveform, as already observed in figure \ref{fig:Pin_R_WT}. The detailed quantitative dependence is determined by $\sum_{n=1}^{\infty} 2 |A_n|^2 \sqrt{n}$, and thus obviously affected by the waveform. The spectral contribution to the power consumption per unit amplitude increases with $n$ as $\sqrt{n}$, so that higher-$n$ modes contribute more to the energetic cost at the same amplitude.

Predicting the turbulent drag reduction rate $R$ is less trivial, since $R$ does not simply derive from the laminar solution (\ref{eq:Generalized_Stokes_layer}) but results from the non-linear interaction between the oscillation of the wall and the near-wall turbulence. Several proposals are available in the literature to link properties of the transverse layer with $R$. In particular, it has been suggested \cite{choi-xu-sung-2002, quadrio-ricco-2004} that $R$ is linked to a parameter $V_R$ that combines a length and an acceleration scale of the spanwise alternating layer. Unfortunately, we have found that this scaling parameter does not work for non-sinusoidal waveforms: when plotting $R$ versus $V_R$, the data for each waveform collapse on straight lines, the slope of which is, however, peculiar to every particular waveform.

In order to identify a universal scaling parameter for the drag reduction rate achieved with different waveforms, the relation between $R$ and the laminar Stokes layer is thus revisited. The Stokes thickness can be interpreted as the maximum wall distance where the oscillation of the wall significantly interferes with the local turbulence, and is typically defined as the largest distance from the wall where the maximum wall-induced spanwise velocity exceeds a threshold velocity $W_{th}$ \cite{quadrio-ricco-2004, touber-leschziner-2012}. For the sinusoidal oscillation the Stokes thickness increases with $W_m$ and scales with $\sqrt{T/Re}$. For the generic waveform, the same dependence holds for every harmonic component. However, the expression (\ref{eq:Generalized_Stokes_layer}) reveals a phase shift among the various harmonic components, so that the correct velocity scale to define the penetration depth cannot simply be $W_m$ anymore. A good candidate for the definition of an effective penetration length for non-sinusoidal oscillations is the mean square value (variance) of the oscillating spanwise velocity, $\overline{ w_{St}(y,t)^2 }$.

A new penetration length, $\ell$, can thus be defined as the distance from the wall where the induced variance of the oscillating velocity drops below a threshold value $\sigma_{th}^2$. From the Stokes solution, this variance for the sinusoid is given by $\overline{w_{St}^2}(y) = {W_m^2}/{2} \, \rme^{-2y / \delta}$, which yields
\begin{equation}
\ell(\sigma_{th}) = \frac{1}{2} \delta \ln \left( \frac{W_m^2}{2 \sigma_{th}^2} \right) .
\label{eq:wth-sigmath-sinusoid}
\end{equation}
For the generic waveform, the variance of the oscillating velocity is given by
\begin{equation}
\overline{w_{St}^2} (y) = W_m^2 \sum_{n=1}^{\infty} 2 |A_n|^2 \rme^{-2 \sqrt{n} y / \delta } ,
\label{eq:Stokes_rms2}
\end{equation}
and this highlights how each mode contributes to the variance with a weighing factor which decays exponentially with increasing $n$, an observation that will be important in the following derivations. We also remark that, in general, the penetration length, $\ell$, cannot be expressed analytically, but must be computed numerically from Eq.~(\ref{eq:Stokes_rms2}). In computing $\ell$, the commonly employed value $W_{th}^+=1.2$ is converted into the equivalent $\sigma_{th}^+=0.85$ which follows form Eq.(\ref{eq:wth-sigmath-sinusoid}) for the sinusoidal waveform.

Figure \ref{fig:prediction} (right) shows that the relationship between $R$ and $\ell$ is indeed similar for all the waveforms, provided the open symbols are not considered. These data points correspond to cases at the largest oscillation periods, where it is known that drag reduction vanishes and the interaction between the streamwise turbulent flow and the slowly oscillating Stokes layer trivially becomes a cyclic reorientation of the former by the latter.
For forcing periods $T^+ < 150$, the right panel of figure \ref{fig:prediction} shows how data for all the waveforms show a linear scaling when $\ell^{3/2}T^{-1/2}$ is used as independent variable, where $T^{-1/2}$ accounts for the physical process of diffusion.
Hence, the turbulent drag reduction rate (at not-too-large oscillation periods) is well predicted by the expression
\begin{equation}
{R = h_1 {\ell^{+}}^{(3/2)} {T^{+}}^{(-1/2)} + h_2}
\label{eq:pred}
\end{equation}
where $h_1$ and $h_2$ are coefficients for which a linear fit of the present data at $Re_\tau=200$ and $T^+<150$ yields $h_1=0.0755$ and $h_2=0.016$.

The strong link between the penetration length $\ell$ and the drag reduction $R$ for $T^+ < 150$ can be further exploited to provide an analytical {\it a priori} estimate of the drag reduction capabilities of a generic waveform, when $W_m$ and $T$ are given. Owing to the $n$ modulation, the sum in (\ref{eq:Stokes_rms2}), under certain constraints discussed later on, can be simply approximated with its first term as
\begin{equation}
\sum_{n=1}^{+\infty} 2 |A_n|^2 \rme^{-2 \sqrt{n} y/\delta} \simeq 2 |A_1|^2 \rme^{- 2 y/\delta}.
\label{eq:approx}
\end{equation}
This approximation allows for an analytical estimate of $\ell$ as
\begin{equation}
{\ell}(\sigma_{th}^2) = \frac{1}{2} \delta \ln \left (\frac{2 W_m^2 |A_1|^2 }{\sigma_{th}^2} \right ) .
\label{eq:lnew}
\end{equation}

In fact, by plugging numerical values of the spectral components in (\ref{eq:Stokes_rms2}), it is easy to realize that for the second Fourier component to produce a contribution to the variance comparable to the first one, it must be $|A_2|/|A_1| \sim 10$. Since for the waveforms considered in the present work the first mode happens to be by far the most energetic, approximation (\ref{eq:approx}) is reasonable. In a waveform where the first mode is not dominant, the spectrum should increase with $n$ at least exponentially; in such a case, this would bring about a dramatic increase of the power consumption and would make such a waveform highly unpractical.

\begin{figure}
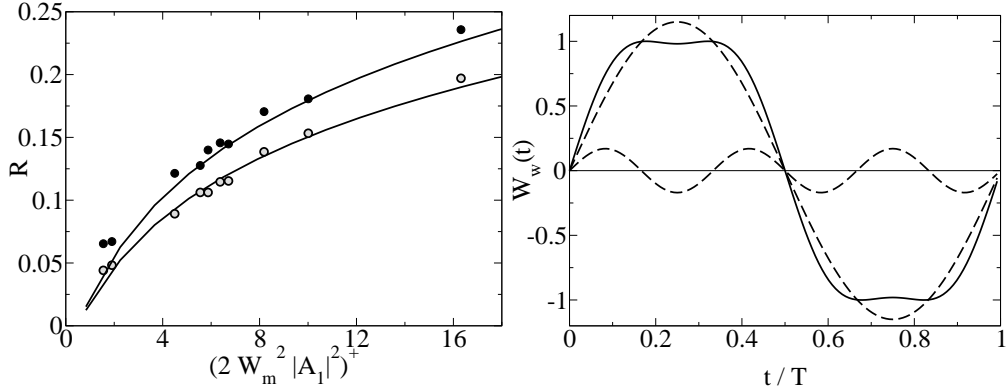

\centering
\includegraphics[width=0.4\textwidth]{figure05a.eps}
\includegraphics[width=0.4\textwidth]{figure05b.eps}
\caption{Left: Turbulent drag reduction rate, $R$, as a function of $2 W_m^2 |A_1|^2$ for $T^+ =125$ (black symbols) and $T^+ =62.5$ (gray symbols). The solid lines are the prediction (\ref{eq:pred}), with $\ell$ computed via the first mode, Eq. (\ref{eq:lnew}). Right: Bi-harmonic waveform (solid line) designed to increase the sinusoidal net energy savings at low $W_m$ values.}
\label{fig:R_scaling}
\end{figure}

In figure \ref{fig:R_scaling} (left), $R$ is plotted against $W_m^2 |A_1|^2$ for $T^+=125$ and $T^+=62.5$, to confirm that the drag reducing properties of the different waveforms are driven by their slowest sinusoidal component, provided its period does not exceed the value $T^+=150$ where there is a change in the nature of the interaction with turbulence. Indeed, when this condition is satisfied, $R$ monotonically increases with $W_m^2 |A_1|^2$. The estimation of $R$ is now possible analytically using Eq.~(\ref{eq:pred}) when $\ell$ is computed through (\ref{eq:lnew}), i.e. using only its first Fourier mode. This estimation of $R$ is included in figure \ref{fig:R_scaling} as a solid line, and shows very good agreement with the data points obtained by DNS. We stress moreover that, within the accuracy of the proposed model, this estimate of $R$ is a lower bound, since any neglected harmonics lead to a further increase of $R$, although in a progressively less significant way.
%
%

Having shown that both $P_{in}$ and $R$ can be predicted purely on the basis of the laminar solution of the Stokes layer, it follows that the net energy saving rate, $S$, the quantity of ultimate interest from the application standpoint, can be predicted too.
Since equations (\ref{eq:Generalized_Pc}) and (\ref{eq:Stokes_rms2}) do not require extensive computations, such predictive capabilities can be used to guide the design of the wall movement when trying to implement an oscillating wall in practice. Lastly, let us remark that the present approach is naturally also valid for the classical sinusoidal oscillation, of which it represents a generalization.

\subsection{Discussion}
\label{sec:discussion}

In the previous sections we have shown that the linearity of the spanwise averaged momentum equation, together with the discovery of a link between drag reduction $R$ and a newly defined penetration depth $\ell$, allows for a simple prediction of the control performances in terms of harmonic components of the waveforms.
Eqns.~(\ref{eq:Generalized_Pc}) and (\ref{eq:Stokes_rms2}) show that the coefficients $W_m A_n$ for $n > 1$, which are non-zero for non-sinusoidal waveforms, affect the power consumption $P_{in}$ more than the variance of the oscillating velocity $\overline{ w_{St}(y,t)^2}$, since each modal contribution to $P_{in}$ is multiplied by $\sqrt{n}$ whereas $\overline{ w_{St}^2}$ is multiplied by $\exp(-\sqrt{n})$.  This implies that non-sinusoidal wall oscillations, at least for $T^+<150$, cannot yield better $S$ than the corresponding sinusoid with the same $W_m A_1$. Moreover, we know that for the sinusoid the point of the parameter space $W_m A_1=W_{m,opt}$ and $T_{opt}$ leads to the largest value of $S$. These two facts explain our observation, put forward in \ref{sec:results}, that at  $W_m=W_{m,opt}$ and $T=T_{opt}$ the sinusoidal waveform is best in terms of $S$. Although our DNS dataset considers a finite set of waveforms only, we have shown that this is a general property.

Along the same reasoning, however, locations in the $(W_m, T)$-space do exist where non-sinusoidal oscillations yield better $S$ than the sinusoid. For example, waveform $(f)$ is found to work better than the sinusoid in the high-$W_m$ regions of the $(W_m,T)$-space. The reason is that, given the maximum wall velocity $W_m$, the amplitude of the base sinusoidal wave $A_1$ for the waveform $(f)$ is very small compared to the sinusoidal one. At the same time the contributions of the higher modes $A_n$ for $n>1$ is rapidly decreasing, since $|A_n|^2 \sim n^{-2}$. This means that this particular waveform has a carrier sinusoidal wave that works near the optimal conditions $(W_{m,opt},T_{opt})$, and the decrease of performances due to the harmonics $A_n$ for $n>1$ is not large enough to worsen the overall net energy saving.

This fact is potentially important for applications where technological constraints might force the wall movement within a suboptimal region of the $(W_m,T)$ space: using non-sinusoidal waveforms with carefully designed energy spectrum might help increasing the control performances. Given the maximum value of $S$ fixed at $(W_{m,opt},T_{opt})$, the distribution of $S$ in the parameter space around this optimum can be improved with non-sinusoidal oscillations. As an example, we designed a waveform aimed at optimizing performance within the reasonable practical constraint of remaining in the low-$W_m$ region. According to the previous arguments, in this region a waveform should be designed where the first mode is larger than the single sinusoid with the same amplitude. By considering only two modes, we have used $4|A_1|^2 = 1.33$ and $4|A_3|^2 = 0.03$. The resulting waveform is the solid line in the right plot of figure \ref{fig:R_scaling}, where the two contributing modes are shown by dashed lines. By means of DNS, it has been found that the newly designed waveform yields a value of $S$ which is indeed increased above that by the sinusoid. Specifically, for $(W_{m,opt}/2,T_{opt})$ we have measured $S=0.05$ for the the sinusoid and $S=0.06$ for this waveform.


\section{Conclusions}
\label{sec:conclusions}

We describe how turbulence control properties, namely the drag reduction and the net energy saving rate, for the spanwise oscillating-wall technique are affected by non-sinusoidal waveforms. Starting from a large but finite set of temporal waveforms, specific trends in the corresponding DNS data are first identified and then generalized thanks to analytical considerations. Based on these findings a model is presented which allows predicting the flow control performance for the generic waveform.

For sinusoidal wall oscillations it is already known that the Stokes layer turns out to be basically insensitive to the background turbulence while it does strongly affect the physics of near-wall turbulence, resulting in significant reduction of the skin-friction drag. In this case a scaling parameter can be defined  that connects properties of the laminar Stokes layer to turbulent drag reduction \cite{choi-xu-sung-2002, quadrio-ricco-2004}.
In the present work, we verify that the laminar solution of the Stokes layer also describes very well the phase- and plane-averaged spanwise velocity profile induced by an arbitrary periodic wall oscillation. The classical scaling parameter for drag reduction is found not to work anymore once
more general oscillating shapes are considered. Hence, we design a new model for the prediction of the drag reduction rate for any forcing waveform including the classical sinusoidal one.


To connect the laminar solution for the Stokes layer to the turbulent drag-reduction properties, a key step is that of re-defining the penetration length $\ell$ of the Stokes layer as the wall distance where the mean squared value of the oscillating velocity decreases below a threshold value. For $T^+<150$, properties of the laminar solution are shown to correlate well with the turbulent drag reduction rate. Moreover, we propose a simple formula that predicts analytically with very good accuracy the global performance of the oscillating wall for the generic temporal waveform, and in particular its net energy saving $S$.

The present analytical considerations show that the sinusoidal waveform gives the overall best net energy saving in the single point $(W_{m,opt}, T_{opt})$ of the parameter space. The developed model provides a new, simple criterion to design a waveform that allows outperforming the sinusoid in locations of the parameter space that differ from $(W_{m,opt}, T_{opt})$. In other words, the present results are not the trivial superposition of well-known results for the sinusoidal components into which a general waveform can be decomposed. Indeed, while the overall power consumption is the sum of the different harmonic contributions, the drag reduction and consequently the net energy saving are not. Although the variance of the Stokes profile depends linearly on the different harmonic contributions, the resulting penetration length $\ell$, which is the scaling parameter that is linked to the turbulent drag reduction rate, is not the simple sum of the various penetration lengths from the different harmonics.

As a concluding note, let us recall how it is now recognized that moving ahead from the temporal oscillation of the wall (Temporal Stokes Layer, TSL) to the space-varying oscillation \cite{viotti-quadrio-luchini-2009} (Spatial Stokes Layer, SSL) and then to the spatio-temporal oscillation, i.e. the streamwise-travelling waves by Quadrio et al \cite{quadrio-ricco-viotti-2009} (Generalized Stokes Layer, GSL) brings about significant improvements of performance, while the underlying physics is very much unaffected. The present work should thus be regarded as the first step towards the long-term aim of building a predictive tool that should be able to deal with the most general case. In this sense, the ultimate goal of our research effort is considering a Generalized Stokes Layer with the further generalization of letting the temporal and spatial waveform free to assume an arbitrary shape.

\begin{acknowledgments}
The authors would like to acknowledge the support of the German Academic Exchange Service (DAAD) for Andrea Cimarelli, of the Japan Society for the Promotion of Science (JSPS) Postdoctoral Fellowship for Research Abroad for Yosuke Hasegawa as well as the support of the German Research Foundation (DFG) through project FR2823/2-1. The infrastructure for carrying out this project was provided by the Center of Smart Interfaces at TU Darmstadt which is greatly acknowledged.
\end{acknowledgments}

\end{document}